\documentclass[aps,pra,reprint,groupedaddress]{revtex4-1}

\usepackage[T1]{fontenc}
\usepackage[utf8]{inputenc}
\usepackage[english]{babel}
\usepackage{bm}	
\usepackage{amsopn}	
\usepackage{amssymb}	
\usepackage{amsmath}	
\usepackage{dsfont}		
\usepackage{braket}		
\usepackage{physics}
\usepackage{graphicx}
\usepackage{svg}
\usepackage{booktabs}	
\usepackage{makecell}	
\usepackage{tabularx}	    

\usepackage{tikz}
\usetikzlibrary{arrows,positioning}
\usepackage{environ}
\usepackage{varwidth}

\begin{document}


\title{Rapid cooling of a strain-coupled oscillator by optical phaseshift measurement}

\author{Signe Seidelin}
\email{signe.seidelin@neel.cnrs.fr}
\affiliation{Univ. Grenoble Alpes, CNRS, Grenoble INP and Institut N\' eel, 38000 Grenoble, France}
\affiliation{Institut Universitaire de France, 103 Boulevard Saint-Michel, 75005 Paris, France}

\author{Yann Le Coq}
\affiliation{LNE-SYRTE, Observatoire de Paris, Universit\' e PSL, CNRS, Sorbonne Universit\' e, Paris, France}

\author{Klaus M{\o}lmer}
\affiliation{Department of Physics and Astronomy, Aarhus University, Ny Munkegade 120, DK-8000 Aarhus C, Denmark}

\author{}
\affiliation{}


\date{\today}

\begin{abstract}
We consider an optical probe that interacts with an ensemble of rare earth ions doping a material in the shape of a cantilever. By optical spectral hole burning, the inhomogeneously broadened transition in the ions is prepared to transmit the probe field within a narrow window, but bending of the cantilever causes strain in the material which shifts the ion resonances. The motion of the cantilever may thus be registered by the phase shift of the probe. By continuously measuring the optical field we induce a rapid reduction of the position and momentum uncertainty of the cantilever. Doing so, the probing extracts entropy and thus effectively cools the thermal state of motion towards a known, conditional oscillatory motion with strongly reduced thermal fluctuations. Moreover, as the optical probe provides a force on the resonator proportional to its intensity, it is possible to exploit the phase shift measurements in order to create an active feedback loop, which eliminates the thermal fluctuations of the resonator. We describe this system theoretically, and provide numerical simulations which demonstrate the rapid reduction in resonator position and momentum uncertainty, as well as the implementation of the active cooling protocol.

\end{abstract}


\maketitle

\section{Introduction}
Mechanical resonators have multiple applications in science and technology and their operation in the quantum regime enable effective coupling to weak perturbations and to a variety of other quantum systems for precision sensing and quantum information processing purposes. The preparation of mechanical oscillators in well-defined quantum states have thus been the target of many efforts, and both thermalization with a low temperature environment~\cite{oconnell2010}, sideband microwave cooling~\cite{teufel2011}, and more elaborate heralding schemes~\cite{Vanner2013,rossi2018} have been employed or theoretically proposed. In this article, we propose and analyze a novel scheme for cooling of a cantilever, which makes use of a rare earth ion ensemble, doped into the cantilever material. In ref.~\cite{Molmer2016} we suggested to prepare such an ensemble by optical hole burning techniques such that light strongly detuned from all the ions can be transmitted through a spectral hole, while bending of the material causes strain and shifts the optical transition frequencies of the ions modifying their dispersive interaction with the light probe. Using realistic parameters, we argued in ref.~\cite{Molmer2016}  that it would be possible to resolve the thermal bending motion of the cantilever in much shorter time than the life time of the spectral hole. Here, we take the analysis further and derive the conditional state of the system subject to continuous homodyne monitoring of the transmitted field.

In Sec.~\ref{physicalsystem}, we introduce our system and motivate a Gaussian ansatz for the state of motion of a vibrational mode of the cantilever and of the quantized probe field. In Sec.~\ref{gaussianstate}, we derive and solve the equations of motion for the first and second moments of the Gaussian phase space distribution of the cantilever motion subject to optical probing. We obtain numerical and approximate analytical expressions for the position and momentum variances and we show sample trajectories for the mean displacement of the cantilever motion, conditioned on realistic measurement records. We argue that the reduced position and momentum uncertainty is equivalent to a cooling of the mechanical motion, and permits definition of an effective temperature of the cantilever far below the surrounding environment. By adding an active feed back mechanism, we show that this low effective temperature can also be translated into an effective ``freezing'' of the resonator where its mean position and momentum are not only precisely known but also constant. Finally, Sec.~\ref{conclusion} provides a brief conclusion and outlook.

\section{The physical system}\label{physicalsystem}

\begin{figure}
  \centering
  \includegraphics[width=8cm]{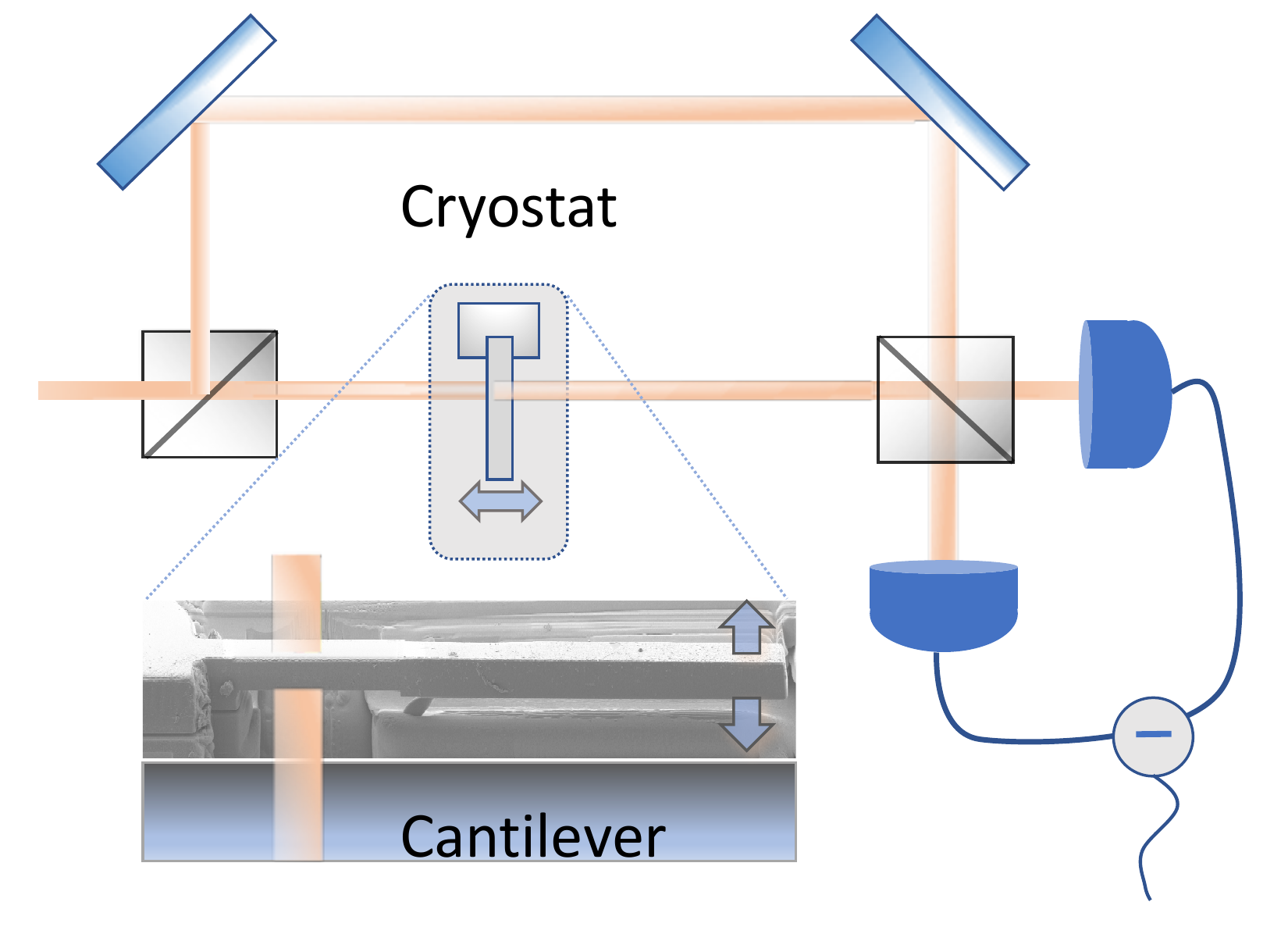}
  \caption{A schematics of the set-up allowing to detect the vibrations of a cantilever using homodyne detection. The dimensions of the cantilever are $100\times10\times10$ $\mu m^3$, for more details, see Sec.~\ref{numericalapplication}.}
  \label{setup}
\end{figure}

We consider the physical system depicted in fig.\ref{setup}, consisting of a transparent cantilever which is probed by a coherent laser beam. The interferometric set-up allows measurement of the quadrature of the transmitted beam which contains information about the bending motion of the cantilever: the cantilever material is doped with rare earth ions featuring a narrow optical transition which is shifted in frequency due to crystal strain. The inhomogeneously broadened absorption profile of the ion dopants is prepared by spectral hole burning to allow transmission of the probe laser field, but when the cantilever is bent, the spectral hole is distorted, and the dispersive interaction with ions causes an optical phase shift. For details, the full protocol of this strain-coupling is outlined in ref.~\cite{Molmer2016}.

The coherent beam of light with flux $\Phi$ can be thought of as a product state of segments of duration $\tau$, each containing a coherent state with average photon number $n=\Phi\tau$. We assume that, prior to interaction with the cantilever, the coherent states have null phase and real argument $\alpha=\sqrt{n}=\sqrt{\Phi \tau}$. In our previous work \cite{Molmer2016} we have shown that, by interacting with a bent cantilever with appropriately prepared spectral hole structure, the light beam experiences a phase shift proportional to the resonator displacement $X_{m}$, namely $\Delta\phi = k X_{m}$. A Fock state $|n\rangle$, for its part, undergoes a quantum phase shift, $|n\rangle \rightarrow e^{-i n \Delta\phi}|n\rangle$, where the phase shift per photon is equal to the phase shift of the coherent field amplitude. We write  $\hat{a}= \alpha+\delta\hat{a}$, such that the number operator $\hat{n}$ can be written
\begin{eqnarray}
\hat{n} & = & \hat{a}^\dagger \hat{a} = (\alpha +\delta\hat{a}^\dagger)(\alpha +\delta\hat{a})\nonumber \\
& \simeq & \alpha^2 + \alpha(\hat{a}-\alpha) +  \alpha(\hat{a}^\dagger-\alpha)  = \alpha(\hat{a}+\hat{a}^\dagger)-\alpha^2\nonumber \\
& = & \sqrt{2}\alpha \hat{X}_{ph} - \alpha^2.
\end{eqnarray}
Utilizing our assumption that the input coherent state $|\alpha\rangle$ have null phase, the effect of the Fock state phase factor $e^{-i n \Delta\phi}$ is hence approximated by the operator $e^{-i\sqrt{2}\alpha \hat{X}_{ph} k \hat{X}_m}=e^{-i\kappa_\tau \hat{X}_{ph} \hat{X}_{m}}$, where $\kappa_{\tau}^2 \equiv \kappa^2\tau \equiv 2 k^2\Phi\tau$. The exponential operator form reflects the unitary evolution of the joint state of the field and mechanical oscillators, which is governed by a coupling Hamiltonian $\hat{H} = \kappa_\tau \hat{X}_{ph} \hat{X}_{m}/\tau$.

\section{Gaussian state formalism}\label{gaussianstate}

We consider the joint quantum state of the mechanical oscillator bending mode and a single incident segment of the probe photon beam. The oscillator Hamiltonian and the interaction between the two systems are second order in their respective position and momentum quadrature operators $(\hat{X}_m,\hat{P}_m,\hat{X}_{ph},\hat{P}_{ph})$, and their time evolution for a short time interval $\tau$ is given by a linear mapping,
\begin{equation} \label{eq:map}
\left(
  \begin{array}{c}
    \hat{X}_m \\
    \hat{P}_m \\
    \hat{X}_{ph} \\
    \hat{P}_{ph} \\
  \end{array}
\right)
\rightarrow
\left(
  \begin{array}{cccc}
    1 & \omega\tau & 0 & 0 \\
    -\omega\tau & 1 & \kappa_\tau & 0 \\
    0 & 0 & 1 & 0 \\
    \kappa_\tau & 0 & 0 & 1 \\
  \end{array}
\right)
\left(
  \begin{array}{c}
    \hat{X}_m \\
    \hat{P}_m \\
    \hat{X}_{ph} \\
    \hat{P}_{ph} \\
  \end{array}
\right).
\end{equation}

The operators are expressed in dimensionless units, such that the variables defining the mechanical resonator are given by $X_m=x_m/x_0$ with $x_0=\sqrt{\hbar/m\omega}$ and $P_m=p_m/p_0$ with $p_0=\sqrt{\hbar m\omega}$.

If we assume an initial thermal state of the cantilever, both systems occupy Gaussian states and while they become correlated, the combined system maintains its Gaussian character due to the interaction. Gaussian states are fully characterized by their first and second order moments, and we shall hence identify the changes in these quantities for the cantilever observables due to the continuous interaction with the field and its subsequent homodyne detection.

For this purpose, we introduce the covariance matrix $\Gamma$ with elements $\Gamma_{ij} = 2 Re(\langle(\hat{q}_i-q_i)(\hat{q}_j-q_j)\rangle$, where $\hat{q_i}$ denotes the four quadrature observables and $q_i$ their expectation values.
Reserving the first two components for the mechanical degrees of freedom and the last two for the field, the covariance matrix separates in blocks,
\begin{equation}
\Gamma=\left(
         \begin{array}{cc}
           {\bf A} & {\bf C} \\
           {\bf C}^T & {\bf B} \\
         \end{array}
       \right)
        \label{Gammamatrix}
\end{equation}
where
\begin{equation}
{\bf A}=\left(
         \begin{array}{cc}
           a_{11} & a_{12} \\
           a_{21} & a_{22} \\
         \end{array}
       \right),
       \label{Amatrix}
\end{equation}
represents the oscillator position and momentum variances and covariances, while the matrix {\bf B} describes the similar quantities for the field and {\bf C} represents correlations between the two systems. Prior to the application of each novel segment of the optical field, which is incident on the mechanical system in a coherent state, $\bf{B}$ and $\bf{C}$ take the initial values

\begin{equation}
{\bf B}_0=\left(
         \begin{array}{cc}
           1 & 0 \\
           0 & 1 \\
         \end{array}
       \right),
{\bf C}_0=\left(
         \begin{array}{cc}
           0 & 0 \\
           0 & 0 \\
         \end{array}
       \right).
       \label{BCmatrices}
\end{equation}

After the interaction of the two systems, the oscillator and the light segment are correlated as described by Eq.(\ref{eq:map}), which transforms the covariance matrix as
\begin{equation} \label{eq:update}
\Gamma \rightarrow {\bf S}\Gamma {\bf S}^T,
\end{equation}
where $\bf{S}$ denotes the 4x4 matrix in Eq.~(\ref{eq:map}).

If the transmitted light segment is discarded after the interaction, we merely retain the upper left block of $\bf{\Gamma}$ as our new mechanical covariance matrix $\bf{A}$, while replacing $\bf{B}$ and $\bf{C}$ by  $\bf{B}_0$ and $\bf{C}_0$ in $\bf{\Gamma}$ in Eq.~(\ref{Gammamatrix}) to accommodate for the interaction with the subsequent coherent segment of the beam. However, rather than discarding the transmitted field, we perform a measurement of the phase rotation of the optical field segment, right after its interaction with the cantilever. This is done by homodyne measurement of the quadrature $P_{ph}$. This measurement yields information about $X_{m}$, and for Gaussian states, the gain in information is represented by the following transformation \cite{Fiurasek2002,Giedke2002,Eisert2003} of the cantilever part of the covariance matrix
\begin{equation}
{\bf A}\rightarrow {\bf A} -\eta {\bf C} \left(
                       \begin{array}{cc}
                         0 & 0 \\
                         0 & 1 \\
                       \end{array}
                     \right) {\bf C}^T,
\end{equation}
where $ {\bf C},\ {\bf C}^T$ are extracted from $\Gamma$ after the update rule Eq.(\ref{eq:update}) and $\eta$ is the detector efficiency.

The field measurement outcome is governed by a Gaussian distribution with mean value $\langle \hat{P}_{ph}\rangle = \kappa_{\tau} \langle \hat{X}_m\rangle$ and variance $1/2$.  Denoting the outcome as $\langle \hat{P}_{ph}\rangle + \chi$, the mean value of the Gaussian distribution of cantilever observables is, indeed, shifted conditioned on the field measurement,
\begin{equation}
\label{eq.Measurement}
\left(
  \begin{array}{c}
    \langle \hat{X}_m\rangle \\
    \langle \hat{P}_m\rangle \\
  \end{array}
\right)
\rightarrow
\left(
  \begin{array}{c}
    \langle \hat{X}_m\rangle \\
    \langle \hat{P}_m\rangle \\
  \end{array}
\right)
+\sqrt{\eta} {\bf C}
\left(
  \begin{array}{c}
    0 \\
    \chi \\
  \end{array}
\right).
\end{equation}

In addition to the deterministic and stochastic evolution of the oscillator covariance matrix and mean values due the field probing and free evolution, we include the equilibration of the oscillator with its thermal environment at rate $\gamma$. This is done by adding the following terms to the rate equations in the continuous limit,
$da_{ii}/dt|_\gamma = -\gamma a_{ii}+\gamma(2\overline{n}+1)$ for $i=1,2$, $da_{ij}/dt|_\gamma=-\gamma a_{ij}$ for $i\ne j$, and $d\langle \hat{X}_m\rangle/dt|_\gamma = -\frac{\gamma}{2}\langle \hat{X}_m\rangle$, and $d\langle \hat{P}_m\rangle/dt|_\gamma = -\frac{\gamma}{2}\langle \hat{P}_m\rangle$. In the absence of any other terms, these rate equations would lead to a steady state with $ a_{11}=a_{22}=2 \rm{Var}(X_m) = 2\rm{Var}(P_m) =2\overline{n}+1$, representing the familiar mean energy  $\frac{1}{2}\langle \hat{X}_m^2 + \hat{P}_m^2\rangle = (\overline{n}+\frac{1}{2})$ of the thermalized oscillator.

To summarize, the mechanical resonator is described by the $2\times 2$ covariance matrix $\bf {A}$ (Eq.~\ref{Amatrix}) that evolves in a deterministic manner, and by mean values $(\langle \hat{X}_m\rangle, \langle \hat{P}_m\rangle)$ that follow from the combination of free evolution and the accumulated stochastic measurement record. This evolution is equivalent to the general quantum trajectory treatment of continuously monitored quantum systems by a stochastic master equation, but it is considerably simplified by the restriction to Gaussian states.

\subsection{Continuous limit}

Our division of the probe beam into segments of duration $\tau$ allows us to take the continuum limit, assuming the derivative to be given by $dx/dt=(x(t+\tau)-x(t))/\tau$. It is a special property of the Gaussian states, that the covariance matrix of the mechanical system evolves in a deterministic manner, independent of the measurement outcome. Putting all terms together, we thus get for the components of the oscillator part of the covariance matrix the following explicit equations:
 \begin{eqnarray}
\frac{da_{11}}{dt} & = & -\eta \kappa^2 a_{11}^2 + \omega(a_{21}+a_{12})-\gamma(a_{11}-(2\overline{n}+1))\nonumber \\
\frac{da_{12}}{dt} & = & - \eta \kappa^2 a_{11} a_{12} - \omega(a_{11}-a_{22})-\gamma a_{12}\nonumber \\
\frac{da_{21}}{dt} & = & - \eta \kappa^2 a_{11} a_{21} - \omega(a_{11}-a_{22})-\gamma a_{21}\nonumber \\
\frac{da_{22}}{dt} & = & \kappa^2 - \eta \kappa^2 a_{12} a_{21} - \omega(a_{21}+a_{12})\nonumber\\
                            &    & -\gamma(a_{22}-(2\overline{n}+1)).
\label{diffeq}
\end{eqnarray}
The first, non-linear term in the equation for $a_{11}$ shows that the variance of $X_m$ is reduced, and the proportionality with the measurement efficiency $\eta$ emphasizes that this squeezing of the oscillator position is conditional on the probing. The first term in the equation for $a_{22}$ shows that the unobserved $P_m$ undergoes an increasing variance due to the interaction with the probe field - a diffusive heating due to the spread in $X_{ph}$ of the incident state. Setting $\eta=0$ corresponds to no detection, and hence absence of the cooling/squeezing effect on $X_m$, while $P_m$ always heats up due to the interaction with the probe field. The effective anti-squeezing of $P_m$ ensures the Heisenberg uncertainty relation remains fulfilled even if there were no free rotation (at $\omega$), no heating and if $X_m$ were probed at unit efficiency. Due to the rotation, however, a mixing of the degrees of freedom subject to squeezing and anti-squeezing and heating leads, ideally, to reduction of both variances. 

Together with the deterministic change of the covariance matrix and mean values of the mechanical position and momentum, the mean values of  $\hat{X}_m$ and $\hat{P}_m$ experience stochastic changes (Eq.~\ref{eq.Measurement}) associated with the measurement outcomes. In the limit of infinitesimal time steps $dt$, the difference $dW$ between the measured value and the expected mean value is stochastic with variance $dW^2 = dt$, corresponding to detector shot noise, and its explicit variation leads to the update equation for the mean values:
\begin{eqnarray}
\langle \hat{X}_m \rangle \rightarrow \langle \hat{X}_m \rangle +\sqrt\eta a_{11} \kappa dW \nonumber \\
\langle \hat{P}_m \rangle \rightarrow \langle \hat{P}_m \rangle +\sqrt\eta a_{21} \kappa dW.
\label{update}
\end{eqnarray}

\subsection{Steady-state solutions}

The nonlinear Eqs.~\ref{diffeq} can be solved analytically in the steady state limit. For $\gamma\ll\omega$ (which is readily fulfilled for realistic parameters, see \ref{numericalapplication}), the rapidly oscillating system is effectively subject to equal strength probing of $X_m$ and $P_m$, and the equations for $a_{11}$ and $a_{22}$ can be replaced by the average of the corresponding equations in Eq.~(\ref{diffeq}). The steady state variances, $\rm{Var}(X_m)= \rm{Var}(P_m)= a_{11}/2$ are then determined from the roots of a single quadratic equation, and we obtain

\begin{equation}
\label{eq:a11_steady_state}
a_{11}=\frac{-\gamma +\{\gamma^2+\eta\kappa^2\big[ \kappa^2+2\gamma(2\bar n+1)\big]\}^{1/2}}{\eta \kappa^2}.
\end{equation}

For the realistic system explored in \ref{numericalapplication}, we furthermore have $\gamma\ll\kappa$, and Eq.~(\ref{eq:a11_steady_state}) further reduces to

\begin{equation}
\label{eq:a11_steady_state_reduced}
a_{11}=\frac{1}{\sqrt{\eta}}\sqrt{1+\frac{2\gamma}{\kappa^2}(2\bar n+1)}.
\end{equation}

This result explicitly reflects the competition between the cooling induced by the  measurements with efficiency $\eta$ and probe interaction strength $\kappa^2$ and the heating with rate $\gamma$. For the physical parameter range of interest, we obtain a significant reduction and favorable square root scaling of the position and momentum variances compared to their values in thermal equilibrium with the environment. We recall, however, that the reduction of these variances does not represent extraction of energy from the oscillator, as the mean position and momentum have finite random values. But since these values are known from the measurement record through Eq.~(\ref{eq:update}), we can either reduce them deterministically by application of a force to the system as demonstrated in Sec.~\ref{numericalapplication}, or we can merely retain our knowledge about their values and subtract them ``in software'' in applications of the system, e.g., for  sensing purposes.

We observe that the Eqs.~(\ref{diffeq},\ref{update}) have the same  formal structure as the Kalman filter equations~\cite{maybeck79}  for the estimated state and the variance of the estimate of a linear dynamical system. This is no coincidence: For a quadratic Hamiltonian with linear evolution of the position and momentum observables, quantum measurement theory assigns a conditional Gaussian quantum state, which fully characterizes the probability distribution for the observables by mean values and a covariance matrix. The evolution of these objects is equivalent to the classical Kalman filter, while the derivation based on quantum theory ensures, e.g., fulfillment of Heisenberg's uncertainty relation.

\section{Numerical investigation and active feedback}\label{numericalapplication}

In order to investigate numerically the outcome of the model in a realistic experimental example, we refer to our previous publication \cite{Molmer2016}. We consider the example of a $100\times10\times10$\,$\mu m^3$ cantilever with a bending mode frequency of $\omega =2\pi\times 1$\,MHz mode and an effective mass $m=1.1\cdot 10^{-11}$\,kg, corresponding to the cantilever depicted in fig.~\ref{setup}. The cantilever material consists of $\rm Y_2SiO_5$ containing a  0.1 \% doping of $\rm Eu^{3+}$ ions, with a $^7F_0 \rightarrow$  $^5D_0$ transition centered at 580 nm and a single ion natural linewidth of $ 2\pi\times122$\,Hz and a measured linewidth of typically $2\pi \times1$\,kHz at a temperature of 3\,K or lower. When using a spectral hole width of 6\,MHz, we previously determined the phase shift $\Delta\phi = k X_m$ to be $0.65 \,\mu$rad for a bending of $x_0=\sqrt{\hbar/m\omega}=1.3$\,fm, corresponding to the width of the quantum ground state of the cantilever. Thus, if $X_m$ is measured in units of $x_0$, $k= 0.65\,\mu$rad. Moreover, assuming a laser intensity of 1\,mW (\emph{i.e.} a photon flux of $\Phi=2.92 \times 10^{-15}$), we have $\kappa^2= 2 k^2\Phi=2\pi\times 197$\,Hz. Furthermore, we assume $\eta=1$, the initial temperature T=400\,mK (corresponding to a bath excitation $\overline{n}=9360$), and the bath coupling $\gamma = 2\pi\times 10$\,Hz.

A first consequence of the application of the probe laser is the shifting of the classical rest position of the resonator. This comes from the action on the mechanical oscillator of the same unitary time evolution operator $e^{-ik\hat{X}_m \hat{n}}$ that yields the phase shift of the field, and it occurs  independently of the measurement back-action. When the probe laser is turned on abruptly from a situation at thermal equilibrium where the resonator position and momentum are centered at zero ($\langle X_m\rangle=\langle P_m\rangle=0$), this leads to a large swing, and the resonator mean values will oscillate for a long time (of the order of $2\pi/\gamma$) before thermalization brings it to the new rest position. It is, however, possible to suppress this swing by applying the laser in a feed-forward procedure as the swing is governed by perfectly deterministic classical equations of motions. In our case, this simply consists in first applying the laser at half power for half a period $ \pi/\omega$ of the resonator oscillation, before applying the laser at full power. Indeed, after the first half period of oscillation, the classical motion arrives at its apex with zero average momentum, and this is the equilibrium phase space position for the oscillator subject to the full laser power.

Second, when the probe laser is on \emph{and} its phase is detected, the continuous monitoring leads to a rapid decrease in the uncertainty in position and momentum of the oscillator. This translates into a rapid decrease in the corresponding variances, and a concomitant localization of the Gaussian state which exhibits oscillations at angular frequency $\omega$ with a random phase and a random amplitude, determined stochastically from the measurement record. Note that because of the constant coupling with the thermal bath, the phase and amplitude will vary in time, but with a very slow rate governed by the coupling coefficient $\gamma$.

Third, from the outcome of the continuous measurement, and the corresponding knowledge of $\langle \hat{X}_m\rangle$ and $\langle \hat{P}_m\rangle$, if the intended application of the system requires so, one can apply a feed-back mechanism which will maintain the resonator position as close as possible to the rest position in phase-space. Owing to the continuous measurement process, the resonator will then rapidly acquire a fixed position with an uncertainty substantially smaller than the one governed by the thermal bath. Because the coupling to the thermal bath is relatively weak, the random fluctuations of the phase and amplitude of the resonator oscillation are relatively slow, and the feed-back mechanism is therefore robust against experimentally unavoidable time delay in the feed-back loop. As a demonstration, we have included time delays up to 1 $\mu$s in this process, without impacting the stability of the servo loop.

\begin{figure}
  \centering
  \includegraphics[width=8.3cm]{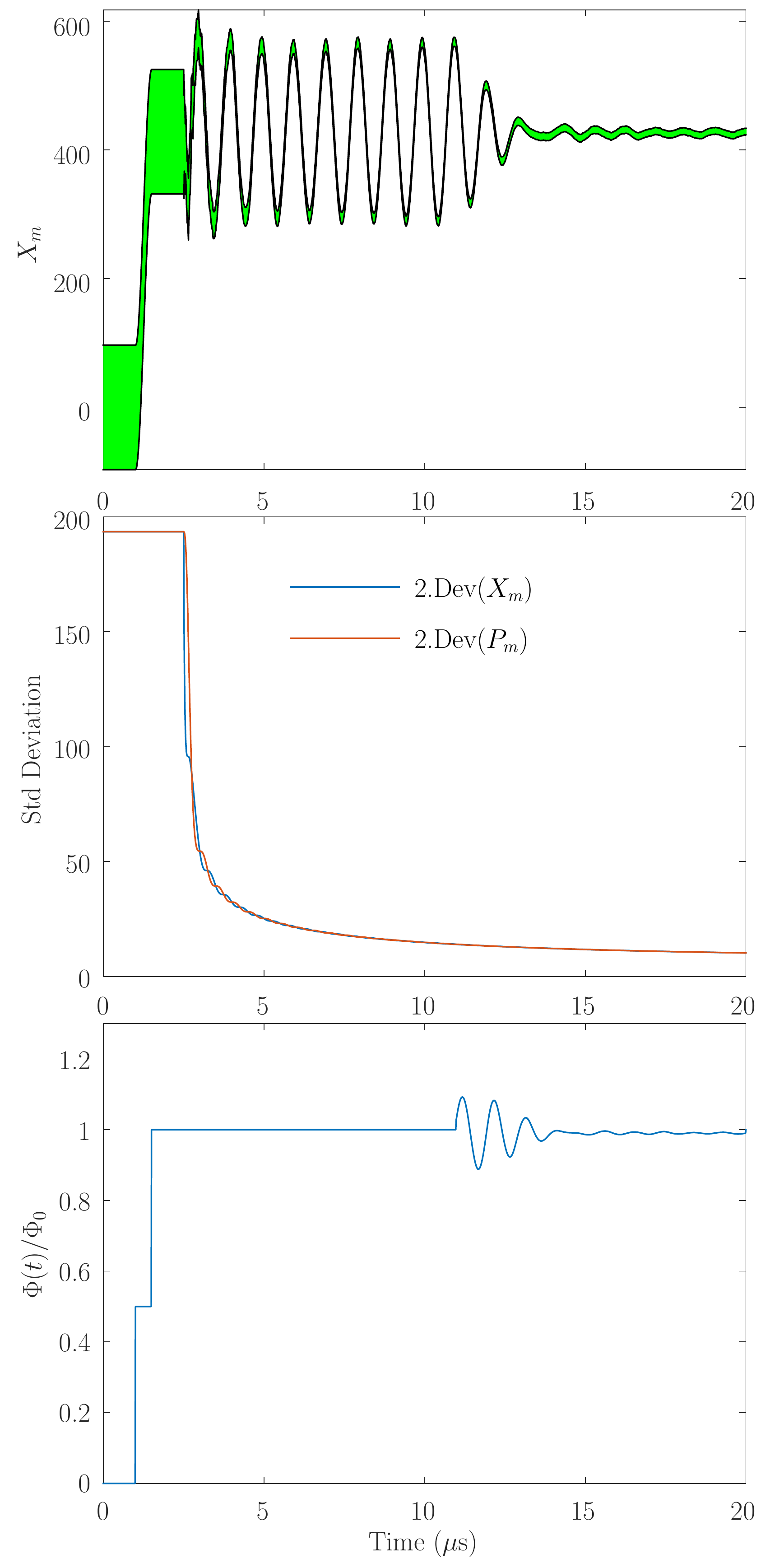}
  \caption{Mechanical resonator subject to optical probing. The physical properties are described in the text ($\omega= 2 \pi \times 1$\,MHz, $\gamma = 2\pi\times 10$\,Hz, laser full power 1\,mW, initial temperature 400\,mK). The sequence used here is the following: 1) we start at thermal equilibrium with no probing laser; 2) after 1\,$\mu$s, we apply the probe laser at half power (feed-forward); 3) at 1.5\,$\mu$s we apply the probe laser at full power; 4) at 2.5\,$\mu$s we detect the probe laser phase continuously, which progressively localizes the resonator on a sine-wave oscillation with random amplitude and phase; 5) after 10 $\mu$s, we apply an active feed-back process (which includes a 1\,$\mu$s delay time) that keeps the resonator near its rest position by acting on the probe laser power. Top panel: mean position of the resonator (the thickness of the curve indicates the uncertainty); middle panel: two times the standard deviation (root-mean-square) of the position and momentum variables; bottom panel: variation in the probe laser power normalized to its final constant value, showing the excitation at half power and the oscillatory feedback modulation, conditioned on the measurement outcome.}
  \label{var} 
\end{figure}

Figure \ref{var} represents the application of this sequence of operation for a resonator with an initial temperature of 400\,mK. Initially, no probe laser is applied and the uncertainty in resonator position is given by its temperature imposed by the thermal bath. When switching on the laser (without detecting its phase), using the feed-forward process described above, we displace the equilibrium position without modifying the effective temperature of the resonator. When the probe laser is applied and its phase after interaction with the cantilever is detected, the oscillator becomes localized in a sine-wave oscillation. After 10 $\mu$s, we apply the active feedback modulation of the probe power to keep the resonator as close as possible to the rest place, and the amplitude of the oscillations rapidly decrease to zero. The feed-back mechanism used here is based on the continuous evaluation of the amplitude and phase of the oscillation at angular frequency $\omega$ and the application of a small modulation of the probe laser intensity at frequency $\omega$, with an amplitude and phase continuously adapted to counter-balance the resonator oscillation, see bottom panel in Fig.~\ref{var}. In our numerical example, we have limited the gain of the feed-back loop to keep the relative modulation of the probe laser power lower than 10 \%, which implies a negligible impact on the measurement back-action process, in order to simplify numerical integration.

Note that the successive application of the above steps is only chosen for clarity. In practice, it is certainly possible to apply the feed-forward ramping of the laser power and implement the phase detection and the feed-back mechanism all together and immediately, so as to accelerate the progress towards the final steady state, should it be desirable for practical applications.

The steady state given in Eq.(\ref{eq:a11_steady_state_reduced}) leads to an uncertainty in the position of the resonator that corresponds to an effective temperature <~0.7~mK (or, equivalently <~15 quanta). This more than 600 fold reduction in effective temperature is a striking demonstration of how effectively the continuous monitoring extracts knowledge of the resonator state. Moreover, a large part of the decrease in $a_{11}$ and $a_{22}$ occurs at the beginning of the process (see fig.~\ref{var}): only 4\,$\mu$s after application of the continuous measurement, the effective temperature is already decreased to $\simeq$ 100\,quanta \emph{i.e.} $\simeq$4\,mK, a 100-fold reduction of the effective temperature. After 50\,$\mu$s, the effective temperature is already within less than 1\% of that of the steady state.

\section{Conclusion}\label{conclusion}


We have in this article presented a Gaussian state formalism that accounts for the evolution of a mechanical oscillator subject to continuous homodyne probing. The measurement outcome is stochastic, and the measurement back action entails a displacement of the oscillator, which one must know to benefit from the significantly reduced variance of the inferred position and momentum of the cantilever.
The increased purity of the quantum state accompanies a reduced entropy, and we refer to the process as measurement induced cooling (a term also used in ref.~\cite{Vanner2013}), as the residual energy of the system is mainly due to a precisely known oscillatory motion in phase space. We also show that this motion that can be arrested by application of a force. In particular, we can use a force which arises from interaction with the probe itself and perform a feedback according to the measurement of the resonator position by varying the intensity of the probe. Future work may incorporate separate studies of the so-called retrodicted state~\cite{Gammelgaard2013,Rossi2018,zhang2017,laverick2019} of the system, in particular, what do we know at time $T$ about the oscillator's position at the earlier time $t$, due to the measurements performed both until $t$ and after $t$, and how this can benefit application of the probed cantilever for force and motional sensing~\cite{tsang2019}.\\

\section*{Acknowledgements}

KM acknowledges support from the Villum Foundation and YLC from the Ville de Paris Emergence Program and from the LABEX Cluster of Excellence FIRST-TF (ANR-10-LABX-48-01), within the Program ``Investissements d'Avenir'' operated by the French National Research Agency (ANR). The project has also received funding from the European Union's Horizon 2020 research and innovation program under grant agreement No 712721 (NanOQTech).


\end{document}